\documentclass[11pt,a4paper,psfrag]{article}

\usepackage{jheppub} 

\usepackage{amssymb}
\usepackage{amsfonts}
\usepackage{latexsym}
\usepackage{pdfsync}

\usepackage[T1]{fontenc} 

\usepackage{epsf,amsfonts}
\usepackage{epsfig}
\usepackage{amsthm}
\usepackage{amsmath}
\usepackage{graphicx}
\usepackage{color}
\usepackage{slashed}
\usepackage{mathrsfs}
\usepackage{float}
\usepackage{color}
\usepackage{esvect}
\usepackage{mathtools}

\definecolor{red}{rgb}{1,0,0}

\title{Chiral Phase Transition with 2+1 quark flavors in an improved soft-wall AdS/QCD Model}

\author[a,b]{Zhen Fang}
\author[b,c]{Yue-Liang Wu}
\author[a,b]{Lin Zhang}

\affiliation[a]{School of Physical Sciences, University of Chinese Academy of Sciences, Beijing 100049, China}
\affiliation[b]{CAS Key Laboratory of Theoretical Physics, Institute of Theoretical Physics, Chinese Academy of Sciences,
Beijing 100190, P. R. China}
\affiliation[c]{International Centre for Theoretical Physics Asia-Pacific (ICTP-AP), University of Chinese Academy of Sciences,
Beijing 100049, China}

\emailAdd{fangzhen@itp.ac.cn}
\emailAdd{ylwu@itp.ac.cn}
\emailAdd{zhanglin@itp.ac.cn}

\abstract{We study the chiral phase transition with $2+1$ quark flavors in an improved soft-wall AdS/QCD model, which can produce the light meson spectrum and many other low-energy quantities consistent with experiments in the two-flavor case. The chiral transition behaviors of the quark condensates at different quark masses are analysed in detail, and the $(m_{u,d},m_{s})$ phase diagram for the quark sector has been obtained from the improved soft-wall model. We find that the features of the calculated phase diagram are completely consistent with the standard scenario, which is supported by lattice simulations and theoretical arguments. The evidence of a tricritical point on the $m_{u,d}=0$ boundary of the $(m_{u,d},m_{s})$ phase diagram is first clearly presented in the bottom-up AdS/QCD.}

\keywords{chiral phase transition, phase diagram, 2+1 flavors, AdS/QCD}

\begin{document}
\maketitle
\flushbottom

\section{Introduction}\label{introduce1}

Study on the phase transition and vacuum structure of quantum chromodynamics (QCD) is very important for our understanding of low-energy physics of strong interaction and even the evolution of our universe \cite{nature-PTD}, which is yet hindered by the nonperturbative nature of low-energy QCD due to the property of asymptotic freedom. In spite of that, much of this field has been investigated over the past decades of studies. It is known that chiral symmetry breaking and quark confinement are two essential features of low-energy QCD. It is widely believed that at high temperature or baryon density QCD will undergo a transition from the hadron matter to the quark-gluon plasma accompanied by the chiral symmetry restoration and color deconfining process. The relation between the chiral and deconfining phase transitions is a hot topic but without final conclusion up to now. The quark-gluon plasma state can now be created in the laboratory, such as the Relativistic Heavy Ion Collider (RHIC) and the Large Hadron Collider (LHC), which also triggers extensive researches on the field of QCD phase transitions although there are still many questions need to be clarified.

To characterize QCD phase transitions, we need to choose proper order parameters. In the chiral limit with zero quark masses, the QCD Lagrangian has an exact chiral symmetry which is spontaneously broken by the QCD vacuum. At high temperatures, this chiral symmetry is expected to be restored, which indicates a chiral phase transition \cite{Gross:1980br}. The proper order parameter to describe the chiral transition is the chiral condensate $\langle{\bar q}q \rangle$ \cite{Nambu:1961tp}. In the pure gauge sector with the quark masses approaching to infinity, the QCD partition function admits a $Z(3)$ symmetry, which is broken at high temperatures due to color screening \cite{McLerran:1981pb}. For the deconfining phase transition, the proper order parameter is the Polyakov loop expectation value $\langle L\rangle$ \cite{Polyakov:1978vu}. In the intermediate region of quark masses, we can still use these order parameters to probe the QCD phase transition behaviors, although there are no exact chiral symmetry and $Z(3)$ symmetry. Besides the observables $\langle{\bar q}q \rangle$ and $\langle L\rangle$, many other quantities can also be used to characterize the transition behaviors, such as the pressure or entropy density in thermal QCD. In this work, we only consider the chiral phase transition with the chiral condensate as order parameter.

\begin{figure}
\centering
\includegraphics[width=75mm,clip=true,keepaspectratio=true]{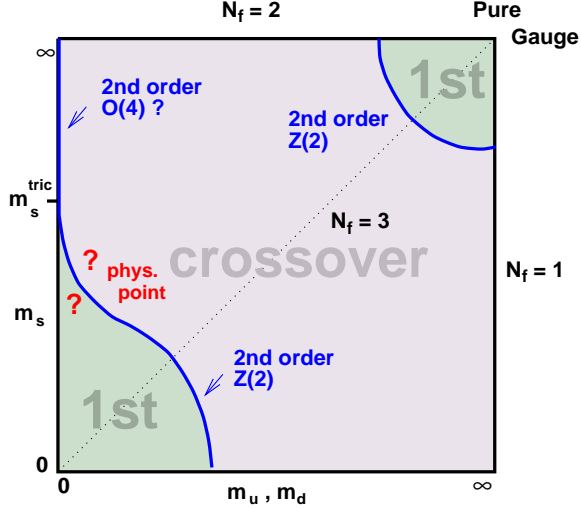}
\caption{The expected phase diagram in the quark mass plane ($m_{u,d}, m_s$) \cite{Laermann:2003cv}.}
\label{columbia-plot}
\end{figure}

The sketched plot in Fig.\ref{columbia-plot} shows the standard scenario of QCD phase diagram in the $(m_{u,d},m_{s})$ quark mass plane \cite{Laermann:2003cv}, which gives the quark mass dependence of the order of phase transition in the $2+1$ flavor case. The lateral axis denotes the up/down quark mass $m_{u,d}$, and the vertical axis denotes the strange quark mass $m_{s}$. The blue curves represent the second-order lines of phase transition, which divide the phase diagram into three parts. The light green regions denote first-order phase transition, while the grey part denotes crossover transition. We can see that the QCD phase transition is first order when quark masses are small, and then goes through a second-order line to a large crossover region with the increasing of quark masses. The up-right corner corresponds to the pure gauge region with very large (or infinite) quark masses, which will not be considered in this work. The diagonal line corresponds to the three-flavor case with $m_{u}=m_{d}=m_{s}$.

We emphasize two points of the phase diagram in Fig.\ref{columbia-plot}. First, there is a tricritical point on the $m_{u,d}=0$ boundary at some finite strange quark mass, which is an intersection with the second-order phase transition line. This tricritical point divides the $m_{u,d}=0$ boundary into two parts. The lower part (smaller $m_{s}$) corresponds to first-order phase transition, while the upper part (larger $m_{s}$) corresponds to second-order transition with an $O(4)$ universal class. It should be noted that the $O(4)$ critical properties have not yet been confirmed definitely. In the case of two light quark flavors (i.e., the upper boundary of the phase diagram), the approximate global chiral symmetry $SU_{L}(2)\times SU_{R}(2)$ indicates that the $O(4)$ scaling behavior might exist in analogy to the $O(4)$ $\sigma$ model. However, if the axial symmetry $U_{A}(1)$ is partially restored at the critical transition temperature, the chiral phase transition might also be first order with the tricritical point disappearing \cite{Pisarski:1983ms,Petreczky:2012rq}. Lattice simulations in recent years indicate that the $U_{A}(1)$ symmetry breaking extends to temperatures around $30\,\mathrm{MeV}$ above $T_{c}$ and support the $O(4)$ scaling behavior of second-order phase transition \cite{Bhattacharya:2014ara}. Second, the transition order around the physical point is not clear in Fig.\ref{columbia-plot}. However, lattice QCD studies indicate that it is an analytic crossover without real singular behaviors of relevant observables \cite{nature-PTD,Bhattacharya:2014ara}.

To study the properties of QCD phase transition, we need to use nonperturbative methods such as lattice QCD, functional renormalization group equations etc., many of which have been developed for several decades and each of them has both merits and shortcomings. For instance, lattice QCD simulations are very hard at small quark masses and are invalid at finite chemical potential due to the sign problem. In recent decades, the holographic approach, viz. the anti-de Sitter/conformal field theory (AdS/CFT) correspondence \cite{Maldacena:1997re,Gubser:1998bc,Witten:1998qj} provides a powerful tool to tackle the low-energy nonperturbative problems of QCD, which is usually called AdS/QCD or holographic QCD. Extensive researches have been born out of this field with the aim to give good descriptions of nonperturbative phenomena with strong interaction \cite{Erlich:2005qh,Karch:2006pv,TB:05,DaRold2005,D3-D7,D4-D6,SS-1,SS-2,Csaki:2006ji,Cherman:2008eh,Dp-Dq,GKK1,GKK2,Sui:2009xe,Sui:2010ay,
Cui:2013xva,Fang:2016nfj,Li:2012ay,Li:2013oda,Shuryak:2004cy,Tannenbaum:2006ch,Policastro:2001yc,Cai:2009zv,Cai:2008ph,Sin:2004yx,Shuryak:2005ia,
Nastase:2005rp,Nakamura:2006ih,Sin:2006pv,Janik:2005zt,Herzog:2006gh,Gubser-drag,Li:2014hja,Li:2014dsa}.

In the holographic framework, many studies have focused on the deconfining phase transition \cite{Herzog:2006ra,BallonBayona:2007vp, Cai:2007zw,Kim:2007em,Andreev-T3,Colangelo:2010pe,Gubser:2008yx,Gubser:2008ny,Gubser:2008sz,Gursoy:2008bu,Gursoy,Gursoy-3,Gursoy:2008za, Finazzo:2014zga,Yaresko:2013tia,Li:2011hp,Cai:2012xh,He:2013qq,Yang:2014bqa,Cui:2014oba}, which can be characterized by different black hole configurations of the dual gravity theory with distinguishable (vanishing or not) Polyakov Loop expectation values. There are also some other works on chiral phase transition in holographic QCD \cite{Colangelo:2011sr,Chelabi:2015cwn,Chelabi:2015gpc,Li:2016smq,Bartz:2016ufc,Bartz:2017jku}. Studies on QCD phase transitions under magnetic fields can be referred to \cite{Evans:2016jzo,Fang:2016cnt,Mamo:2016xco,Dudal:2016joz,Dudal:2018rki, Ballon-Bayona:2017dvv}. As the chiral phase transition is related to the chiral dynamics of QCD, a satisfying holographic description of chiral transition should combine with other important low-energy properties of hadron physics, in which the chiral symmetry breaking plays a dominant role. In this work, we consider the $2+1$ flavor chiral phase transition in an improved soft-wall AdS/QCD model \cite{Fang:2016nfj}, which has been shown to reproduce the light meson spectrum and many other low-energy quantities consistent with experiments, and also the right chiral transition behavior in the two-flavor case. The magnetic effects have also been studied in this model \cite{Fang:2016cnt}, which displays inverse magnetic catalysis in the chiral transition indicated by lattice simulations \cite{Bali:2011qj,Bali:2012zg}.

The paper will be organized as follows. In sec.\ref{2flavorcase1}, we give a brief introduction of the improved soft-wall AdS/QCD model with two flavors \cite{Fang:2016nfj}. In sec.\ref{2plus1generalize1}, we generalize the improved soft-wall model with two flavors to the case of $2+1$ flavors. In sec.\ref{Chiphasetransit}, we study the chiral phase transition for $2+1$ quark flavors in the improved soft-wall model and the ($m_{u,d}, m_s$) phase diagram will be obtained and compared with that shown in Fig.\ref{columbia-plot}. In sec.\ref{gamma-depend}, we study the parameter dependence of the $2+1$ flavor chiral transition and the condition in which the standard scenario of the phase diagram exists. In sec.\ref{conclusion}, we conclude our work and give a brief discussion of the chiral phase transition in the improved soft-wall model.

\section{The improved soft-wall AdS/QCD model with two flavors}\label{2flavorcase1}

We use the pure $\mathrm{AdS}_5$ spacetime as the bulk background with the metric ansatz:
\begin{equation}\label{metric}
ds^2=e^{2A(z)}\left(\eta_{\mu\nu}dx^{\mu}dx^{\nu}-dz^2\right) \,,
\end{equation}
where $A(z)=-\mathrm{log}\frac{z}{L}$ is scaled by the $\rm{AdS}$ curvature radius $L$ (dropped below for simplicity), and the four-dimension (4D) metric convention has been chosen as $\eta_{\mu\nu}=(+1,-1,-1,-1)$.

The action of the improved soft-wall AdS/QCD model in the two-flavor case \cite{Fang:2016nfj} can be written as
\begin{equation}\label{meson-action}
S_M=\int d^{5}x\,\sqrt{g}\,e^{-\Phi(z)}\,{\mathrm{Tr}}\{|DX|^{2}-m_5^2(z)|X|^{2}
-\lambda |X|^{4}-\frac{1}{4g_{5}^2}(F_{L}^2+F_{R}^2)\},
\end{equation}
where $D^MX=\partial^MX-i A_L^MX+i X A_R^M$ and $F_{L,R}^{MN}=\partial^MA_{L,R}^N-\partial^NA_{L,R}^M-i[A_{L,R}^M,A_{L,R}^N]$ with
$A_L^M=A_L^{a,M}t_L^a$, $A_R^M=A_R^{a,M}t_R^a$, in which $t_L^a$ and $t_R^a $ are the generators of $\mathrm{SU}(2)_L$ and $\mathrm{SU}(2)_R$. The dilaton field $\Phi(z)=\mu_g^2\,z^2$ leads to the Regge behavior of meson spectrum \cite{Karch:2006pv}. The key point of this model is the introduction of a quartic term of the bulk scalar field $X(x,z)$ and a $z$-dependent bulk mass $m_5^2(z)$ of $X(x,z)$, which are closely related to the chiral symmetry breaking and thus are crucial for a consistent description of both meson spectrum and chiral phase transition \cite{Fang:2016nfj}. As pointed out in \cite{Fang:2016nfj}, the $z$-dependence of the bulk scalar mass might originate from the running quark mass anomalous dimension which can be linked with $m_5^2$ by the AdS/CFT dictionary $m_5^2=(\Delta-p)(\Delta+p-4)$ with $\Delta$ the dimension of the $p$-form operator. The UV and IR asymptotic behaviors of $m_5^2(z)$ are well constrained by the equation of motion (EOM) of the vacuum expectation value (VEV) of $X(x,z)$ and the mass split of the chiral partners at highly excited states. In \cite{Fang:2016nfj}, we have used the simplest form $m_5^2(z)=-3-\mu_c^2\,z^2$ in accordance with the constrained UV and IR asymptotics.

The mass spectra of $\pi$, $f_{0}$, $\rho$ and $a_{1}$ mesons have been calculated after fixing all the parameters of the model in \cite{Fang:2016nfj}. We only list some of the results in Table.\ref{2f-mesonspectra}. Other low-energy quantities such as the pion form factor and the decay constants of $\pi$, $\rho$, $a_{1}$ mesons have also been calculated in the improved soft wall model, and the results are consistent with the experimental values. Further more, this model also reproduces the same two-flavor chiral transition behavior as that shown in Fig.\ref{columbia-plot}. In the chiral limit, it gives a second-order chiral phase transition, and at finite quark masses it displays a crossover behavior. Hence, it makes sense to generalize this model to include the strange quark sector. Next, we will consider the chiral phase transition for $2+1$ flavors in the framework of the improved soft-wall AdS/QCD model.
\begin{table}\label{2f-mesonspectra}
\begin{center}
    \begin{tabular}{ccccccc}
       \hline\hline
        (MeV) & 0 & 1 & 2 & 3 & 4 & 5 \\
       \hline\hline
       $\pi_{\mathrm{exp}}$  & 139.6 & $1300\pm100$ & $1812\pm12$ & $2070\pm35$ & $2360\pm25$ & ---  \\
       \hline
       $\pi_{\mathrm{ads}}$  & 139.6 & 1296 & 1753 & 2051 & 2277 & 2467  \\
       \hline
       $f_{0\mathrm{exp}}$ & $400-550$ & $1200-1500$ & $1722^{+6}_{-5}$ & $1992 \pm 16$ & $2189 \pm 13$ & --- \\
       \hline
       $f_{0\mathrm{ads}}$ & 586 & 1346 & 1743 & 2016 & 2232 & 2420  \\
       \hline
       $\rho_{\mathrm{exp}}$  & $775.26\pm0.25$ & $1465\pm25$ & $1570\pm 36$ & $1720\pm 20$ & $1909\pm17$ & $2150\pm40$  \\
       \hline
       $\rho_{\mathrm{ads}}$  & 880 & 1245 & 1524 & 1760 & 1968 & 2156  \\
        \hline
       $a_{1\mathrm{exp}}$  & $1230\pm40$ & $1647\pm22$ & $1930^{+30}_{-70}$ & $2096\pm17$ & $2270^{+55}_{-40}$ & ---  \\
       \hline
       $a_{1\mathrm{ads}}$  & 1121 & 1608 & 1922 & 2156 & 2352 & 2526  \\
       \hline\hline
     \end{tabular}
\caption{The meson spectra obtained from the improved soft-wall AdS/QCD model with two flavors \cite{Fang:2016nfj}. The parameters in the calculation are fixed to be $m_{q}=3.366$ MeV, $\mu_g=440$ MeV, $\mu_c=1180$ MeV, $\lambda=33.6$. The experimental data are taken from \cite{Agashe:2014kda}.}
\end{center}
\end{table}

\section{Generalization to the $2+1$ flavor case}\label{2plus1generalize1}

\subsection{General setup}\label{setup1}

It has been shown that an additional term, viz. the 't Hooft determinant $\mathrm{det}[X]$ needs to be considered for the correct realization of chiral phase transition in the three-flavor case \cite{Chelabi:2015cwn,Chelabi:2015gpc}. We drop the sector of chiral gauge fields in the action (\ref{meson-action}), which is irrelevant for chiral phase transition in our consideration. Thus the action to be addressed in the 2+1 flavor case is
\begin{equation}\label{2+1f-action}
S=\int d^{5}x\,\sqrt{-g}\,e^{-\Phi(z)}\,{\mathrm{Tr}}\{|D X|^{2}+m_5^2(z)|X|^{2} +\lambda\,|X|^{4} +\gamma\,\mathrm{det}[X]\} \,.
\end{equation}

To study the chiral transition at finite temperatures in the holographic framework, we use the AdS-Schwarzchild black hole as the simplest ansatz
\begin{equation}\label{BH-metric}
ds^2=e^{2A(z)}\left(f(z)dt^2-dx^{i\,2}-\frac{dz^2}{f(z)}\right)
\end{equation}
with
\begin{equation}
f(z)=1-\frac{z^4}{z_h^4} \,,
\end{equation}
where the horizon of the black hole $z_h$ is related to the Hawking temperature $T$ by the formula
\begin{equation}\label{hawking-T}
T=\frac{1}{4\pi}\left|\frac{df}{dz}\right|_{z_{h}}=\frac{1}{\pi z_{h}}.
\end{equation}
As the close relations with the properties of linear confinement and chiral symmetry breaking, the dilaton field $\Phi(z)$ and the bulk scalar mass $m_5^2(z)$ will take the same forms as those in the two-flavor case
\begin{equation}\label{Phi-m5}
\Phi(z)=\mu_g^2\,z^2 \,, \qquad  m_5^2(z)=-3-\mu_c^2\,z^2 \,.
\end{equation}

Following the previous studies in \cite{Chelabi:2015cwn,Chelabi:2015gpc}, the VEV of the bulk scalar field $X(z)$ in the $2+1$ flavor case is assumed to be
\begin{equation}
\label{VEVs}
\langle X \rangle=\frac{1}{\sqrt{2}}
\begin{pmatrix}
\chi_u(z) & 0 & 0\\
0 & \chi_d(z) & 0 \\
0 & 0 & \chi_s(z)
\end{pmatrix}
\end{equation}
with $\chi_u(z) = \chi_d(z)$. According to AdS/CFT, the chiral condensates of $u\,(d)$ and $s$ quarks are contained in the UV expansion of $\langle X\rangle$ \cite{DaRold2005,Erlich:2005qh}, thus to study the chiral transition behavior we only need to consider the EOM of $\langle X\rangle$, which can be derived from the action (\ref{2+1f-action}) as
\begin{align}
& \chi_{u}'' +\left(\frac{f'}{f}+3A'-\Phi'\right)\chi'_{u} -\frac{e^{2A}}{f}\left(m_5^2\chi_u +\lambda\chi_u^3 +\frac{\gamma}{2\sqrt{2}}\chi_u\chi_s \right) =0 \,, \label{vevX-eom1} \\
& \chi''_{s} +\left(\frac{f'}{f}+3A'-\Phi'\right)\chi'_{s} -\frac{e^{2A}}{f}\left(m_5^2\chi_s +\lambda\chi_s^3 +\frac{\gamma}{2\sqrt{2}}\chi_{u}^{2}\right) =0 \,.  \label{vevX-eom2}
\end{align}

\subsection{Boundary conditions}\label{bc-EOM}

By the prescription of AdS/CFT \cite{DaRold2005,Erlich:2005qh}, the UV asymptotic forms of $\chi_u(z)$ and $\chi_s(z)$ can be directly solved from Eqs. (\ref{vevX-eom1}) and (\ref{vevX-eom2})
\begin{align}
\chi_u(z \sim 0) &=m_u\,\zeta\,z-\frac{m_u\, m_s\,\gamma \, \zeta^2}{2\sqrt2}z^2+\frac{\sigma_u}{\zeta}z^3+\frac{1}{16}m_u\,\zeta\left(-m_s^2 \, \gamma^2\,\zeta^2-m_u^2\, \gamma^2 \,\zeta^2 \right.\nonumber\\
&\,\quad \left. +8 m_u^2 \,\zeta^2 \,\lambda+16 \,\mu_g^2-8 \,\mu_c^2\right)z^3 \,\mathrm{ln}z +\cdots \,, \label{asyExpOfChius1}  \\
\chi_s(z \sim 0) &=m_s\,\zeta\,z-\frac{m_u^2\,\gamma \, \zeta^2}{2\sqrt2}z^2+\frac{\sigma_s}{\zeta}z^3+\frac{1}{8}\left(-m_s\,m_u^2 \, \gamma^2\,\zeta^3+4\,m_s^3\,\zeta^3\,\lambda \right. \nonumber \\
&\,\quad \left. +8 m_s \,\zeta \,\mu_g^2-4 \,m_s\,\zeta\,\mu_c^2 \right)z^3 \,\mathrm{ln}z +\cdots \,, \label{asyExpOfChius2}
\end{align}
where $m_u$, $m_{s}$ denote the quark masses, and $\sigma_u$, $\sigma_s$ are the chiral condensates. The normalization constant $\zeta=\frac{\sqrt{N_c}}{2\pi}$ is fixed by the correct $N_c$ scaling behavior of the quark mass and chiral condensate \cite{Cherman:2008eh}. Note that there are only two independent parameters in the UV expansion of $\chi_u(z)$ or $\chi_s(z)$, i.e., the quark mass $m_{u,s}$ and the chiral condensate $\sigma_{u,s}$, which are most relevant in our discussion of chiral phase transition.

As Eqs. (\ref{vevX-eom1}) and (\ref{vevX-eom2}) are a system of second-order nonlinear ordinary differential equations, we must impose proper boundary conditions to solve them. The first derivatives of $\chi_u(z)$ and $\chi_s(z)$ at $z=0$ will be used as the UV boundary condition in the numerical calculation
\begin{align}\label{boundCond1}
\chi_u^{\prime}(0) =m_u\,\zeta \,, \qquad  \chi_s^{\prime}(0) =m_s\,\zeta \,.
\end{align}
Note that Eqs. (\ref{vevX-eom1}) and (\ref{vevX-eom2}) are singular at the horizon of the black hole, which can be seen from the expansion of these equations at $z=z_h$,
\begin{align}
&\frac{-12\,\chi_u(z_h)-4\,z_h^2\,\mu_c^2\,\chi_u(z_h)+\sqrt{2}\,\gamma\,\chi_s(z_h)\,\chi_u(z_h)+4\,\lambda\,\chi_u^3(z_h)
+16\,z_h\,\chi_u'(z_h)}{16\,z_h\,(z-z_h)}+\mathcal{O}(z-z_h),\\
&\frac{-12\,\chi_s(z_h)-4\,z_h^2\,\mu_c^2\,\chi_s(z_h)+\sqrt{2}\,\gamma\,\chi_u^2(z_h)+4\,\lambda\,\chi_s^3(z_h)
+16\,z_h\,\chi_s'(z_h)}{16\,z_h\,(z-z_h)}+\mathcal{O}(z-z_h).
\end{align}
These singular forms at horizon $z=z_h$ supply us with a natural IR boundary condition to guarantee the regular near-horizon behaviors of $\chi_u(z)$ and $\chi_s(z)$, i.e.,
\begin{align}
& -12\,\chi_u(z_h)-4\,z_h^2\,\mu_c^2\,\chi_u(z_h)+\sqrt{2}\,\gamma\,\chi_s(z_h)\,\chi_u(z_h)+4\,\lambda\,\chi_u^3(z_h)+16\,z_h\,\chi_u'(z_h) =0, \label{boundCond2-1} \\
& -12\,\chi_s(z_h)-4\,z_h^2\,\mu_c^2\,\chi_s(z_h)+\sqrt{2}\,\gamma\,\chi_u^2(z_h)+4\,\lambda\,\chi_s^3(z_h)+16\,z_h\,\chi_s'(z_h) =0. \label{boundCond2-2}
\end{align}

From the Eqs. (\ref{vevX-eom1}), (\ref{vevX-eom2}) and with the boundary conditions (\ref{boundCond1}), (\ref{boundCond2-1}) and (\ref{boundCond2-2}), we can solve the VEVs of the bulk scalar field $\chi_u(z)$ and $\chi_s(z)$ numerically by the spectral collocation method \cite{Shen:2011,Yin:2013}. For one set of quark masses ($m_{u}, m_s$) and a given value of temperature, there will be one set of (or several sets of) solutions ($\chi_{u},\chi_{s}$) with regular IR behaviors at the black hole horizon $z_{h}$. Using the UV asymptotic expansions (\ref{asyExpOfChius1}) and (\ref{asyExpOfChius2}), we can extract the values of the condensates ($\sigma_{u},\sigma_{s}$) from the solutions ($\chi_{u},\chi_{s}$).

\section{Chiral phase transition in the $2+1$ flavor case}\label{Chiphasetransit}

\subsection{Parameters and the main results}\label{parameters1}

After fixing the general setup, we now study the $2+1$ flavor chiral phase transition in our model. To solve the Eqs. (\ref{vevX-eom1}) and (\ref{vevX-eom2}), we need first to fix the four parameters $\mu_{c}$, $\mu_{g}$, $\gamma$ and $\lambda$ in the model. It should be noted that the parameters determined by the hadron spectra in the two-flavor case \cite{Fang:2016nfj} have little reference value in our case here as the strange quark contributions cannot be neglected in the $2+1$ flavor case. Thus, to fix the parameters of the model, we need to calculate the low-energy quantities with the ingredients related to strange quark included, which will not be addressed in this work. Nevertheless, as pointed out in \cite{Fang:2016nfj}, the parameters $\mu_{c}$ and $\mu_{g}$ are intimately linked with the properties of chiral symmetry breaking and linear confinement. The value of $\mu_{c}$ is around the energy scale of chiral symmetry breaking $\mu_c\sim \Lambda_{\chi} \sim 1~\mathrm{GeV}$, while the value of $\mu_{g}$ is close to the $\Lambda_{\mathrm{QCD}}$ energy scale $\mu_g \sim \Lambda_{\mathrm{QCD}}$ \cite{Shuryak:1988ck}. These two parameters have intrinsic meanings and should not be changed significantly. Here we just take the same values of $\mu_{c}$ and $\mu_{g}$ as in \cite{Fang:2016nfj}, i.e., $\mu_{c}=1180\mathrm{MeV}$ and $\mu_{g}=440\mathrm{MeV}$. The choosing of the couplings $\gamma$ and $\lambda$ is relatively arbitrary, except that $\gamma$ must be negative to generate first-order phase transition \cite{Chelabi:2015gpc}, and we take $\gamma=-22.6$ and $\lambda=16.8$ in our analysis below.

We first sketch our main results, which are summarized in the $(m_{u,d},m_{s})$ phase diagram in Fig.\ref{ads-phase-diagram1}. The detailed analysis in different cases will be given below. Note that we do not address the pure gauge sector corresponding to the up-right corner of the phase diagram. In Fig.\ref{ads-phase-diagram1}, the blue curve represents the second-order transition line, which divides the $(m_{u,d},m_{s})$ plane into a first-order transition region and a crossover region (note the different coordinate range of the $m_{s}$ and $m_{u,d}$ axes). We can see that the calculated phase diagram is fully consistent with the phase diagram shown in Fig.\ref{columbia-plot}, at least qualitatively. We also see that the physical point is in the crossover region with parameter values given above. The most interesting thing is that a tricritical point also appears on the $m_{u}=0$ boundary of the calculated phase diagram, which supports the standard scenario in Fig.\ref{columbia-plot}.
\begin{figure}
\centering
\includegraphics[width=80mm,clip=true,keepaspectratio=true]{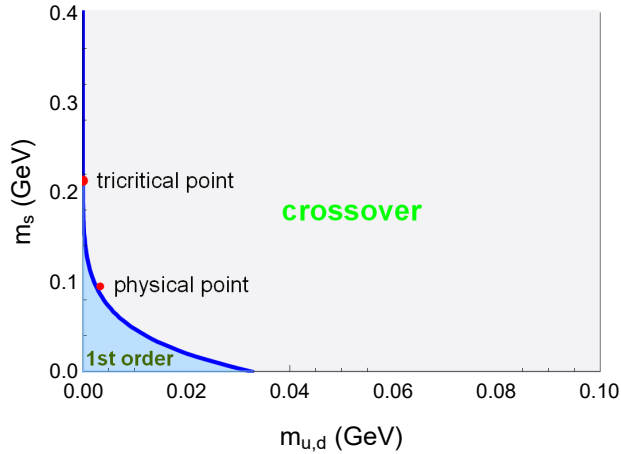}
\caption{The calculated ($m_{u,d}, m_s$) phase diagram in the improved soft-wall model, where the parameters are taken to be $\mu_{c}=1180\mathrm{MeV}$, $\mu_{g}=440\mathrm{MeV}$, $\gamma=-22.6$ and $\lambda=16.8$. The blue curve denotes the second-order transition line.}
\label{ads-phase-diagram1}
\end{figure}

\subsection{The physical point and a typical first-order transition point}\label{physicalpoint1}

We proceed to give the detailed analysis of the calculated phase diagram in Fig.\ref{ads-phase-diagram1}. According to sec.\ref{bc-EOM}, provided that the quark masses ($m_{u},m_{s}$) are given, we can solve the VEVs ($\chi_{u},\chi_{s}$) from the Eqs. (\ref{vevX-eom1}), (\ref{vevX-eom2}) at each temperature and extract the chiral condensates ($\sigma_{u},\sigma_{s}$) from the UV expansion of ($\chi_{u},\chi_{s}$), and then obtain the chiral transition properties of ($\sigma_{u},\sigma_{s}$).

\begin{figure}[h]
\begin{center}
\includegraphics[width=68mm,clip=true,keepaspectratio=true]{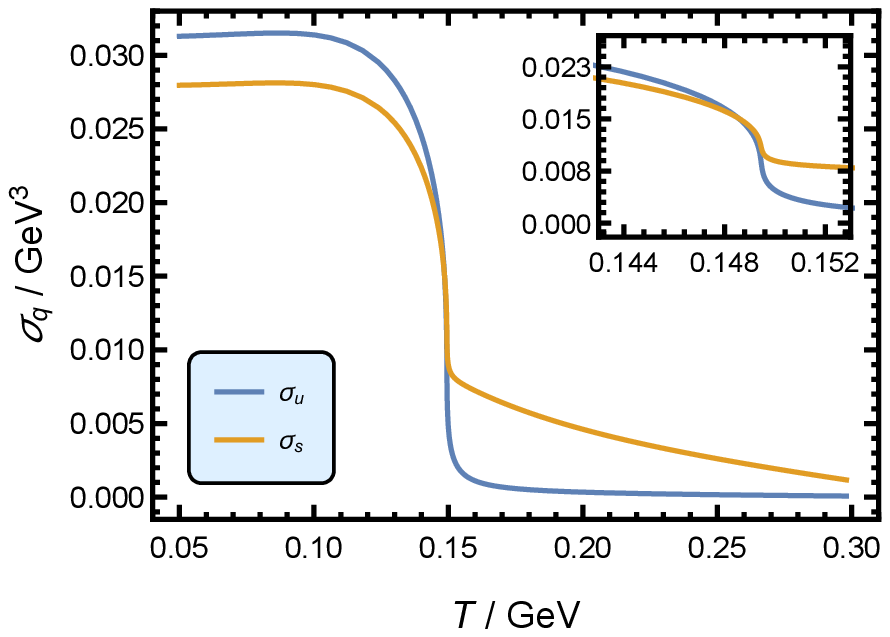}
\hspace*{0.6cm}
\includegraphics[width=65mm,clip=true,keepaspectratio=true]{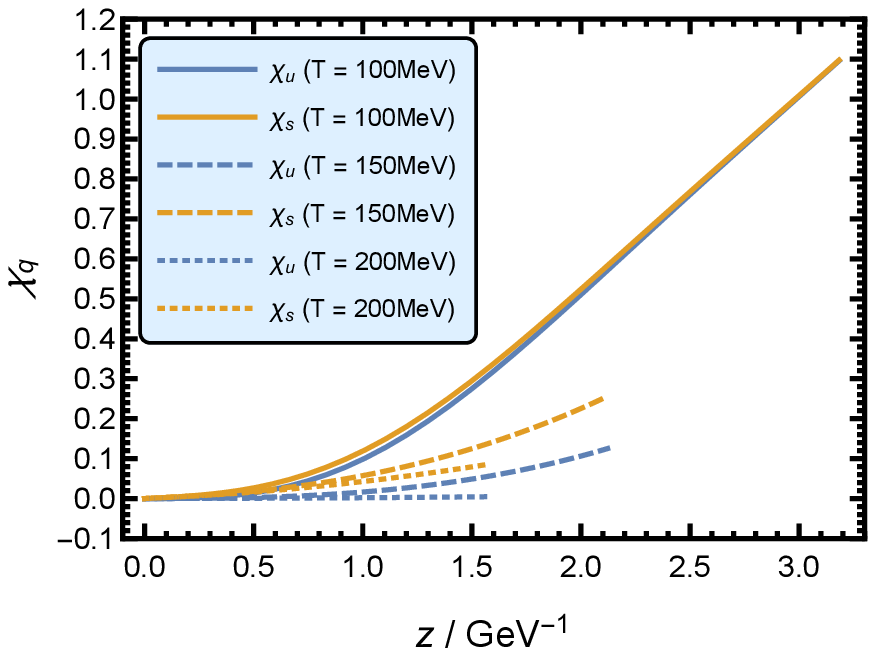} \vskip -1cm \hskip 5.5 cm
\end{center}
\caption{Left: the chiral transition behavior of $\sigma_u$ and $\sigma_s$ with temperature $T$ at the physical point $m_{u}=3.336\,\mathrm{MeV}$, $m_{s}=95\,\mathrm{MeV}$. Right: the solutions of VEVs ($\chi_{u},\chi_{s}$) at temperatures $T=100, 150, 200\,\mathrm{MeV}$.}
\label{chiralPhaseTransitionAndSol1-1}
\end{figure}

We present the calculated results at the physical point with $m_{u}=3.336\,\mathrm{MeV}$ and $m_{s}=95\,\mathrm{MeV}$ \cite{Agashe:2014kda} in Fig.\ref{chiralPhaseTransitionAndSol1-1}.  The left panel shows the chiral transition of the condensates $\sigma_u$, $\sigma_s$ with temperature $T$, from which we can see the crossover behavior obviously. The right panel shows the solutions of VEVs ($\chi_{u},\chi_{s}$) at three different temperatures. We can see that ($\chi_{u},\chi_{s}$) would approach to the zero solution (0,0) with the increasing of temperature $T$, which is easy to understand. As the temperature characterizes the energy scale, the physical quark masses can be neglected at very high temperatures, in which case the Eqs. (\ref{vevX-eom1}), (\ref{vevX-eom2}) have null solution $(\chi_{u},\chi_{s})=(0,0)$. From the left panel of Fig.\ref{chiralPhaseTransitionAndSol1-1} we also see that at low temperatures the condensate $\sigma_{s}$ is suppressed compared with the condensate $\sigma_{u}$, however, after crossing the pseudo-critical transition region around $T\simeq 0.15\, \mathrm{GeV}$, $\sigma_{u}$ drops more quickly than $\sigma_{s}$ and becomes smaller than $\sigma_{s}$ at larger temperatures.

\begin{figure}[h]
\begin{center}
\includegraphics[width=68mm,clip=true,keepaspectratio=true]{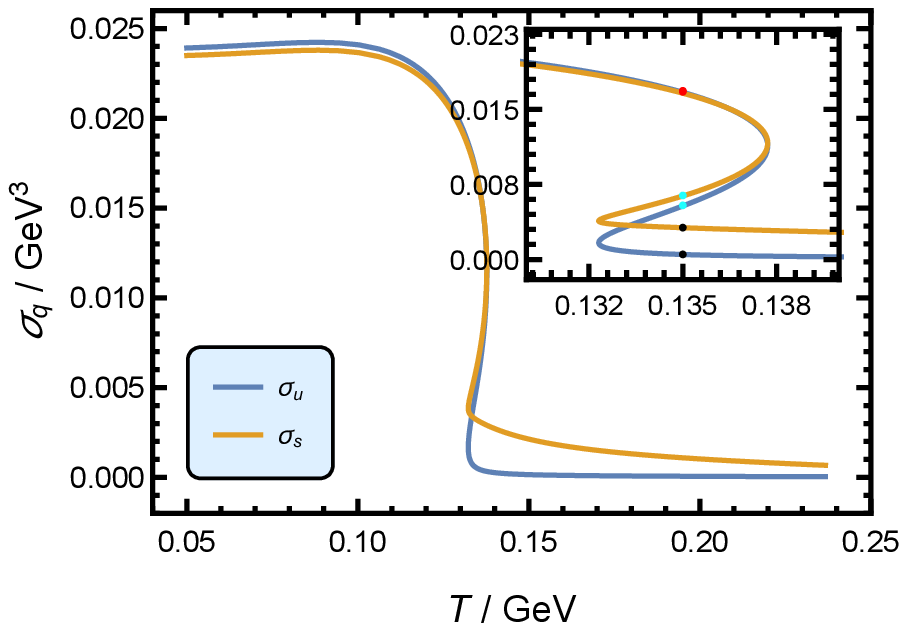}
\hspace*{0.6cm}
\includegraphics[width=65mm,clip=true,keepaspectratio=true]{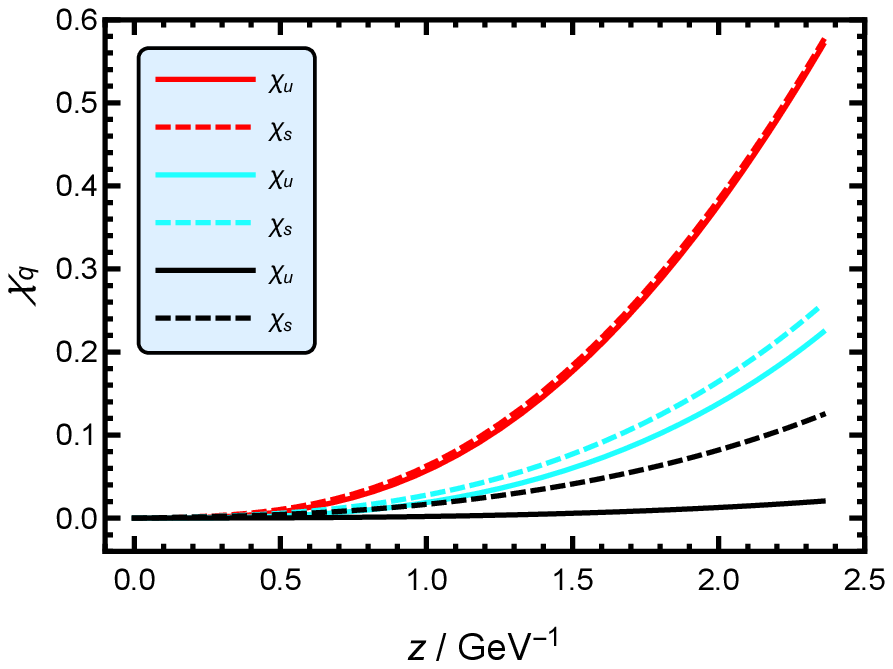} \vskip -1cm \hskip 5.5 cm
\end{center}
\caption{Left: the chiral transition behavior of $\sigma_u$ and $\sigma_s$ with temperature $T$ at the point $(m_{u},m_{s})=(1\,\mathrm{MeV},20\, \mathrm{MeV})$. Right: the three solutions of the VEVs ($\chi_{u},\chi_{s}$) at $T=135\,\mathrm{MeV}$, where the corresponding condensates ($\sigma_u,\sigma_s$) in the left panel (see the insert) have been dotted with the same color as that of the solutions in the right panel.}
\label{firstorder-PT1}
\end{figure}

In Fig.\ref{firstorder-PT1}, we also display a typical first-order phase transition at the point $(m_{u},m_{s})=(1\,\mathrm{MeV},20\,\mathrm{MeV})$.
The left panel shows the chiral transition behavior of the condensates ($\sigma_u$, $\sigma_s$) with the temperature T, from which we can see an inflection in the transition region with a sudden decreasing of the condensate (see the insert), which indicates a first-order phase transition. Note that the Eqs. (\ref{vevX-eom1}) and (\ref{vevX-eom2}) have three solutions at each temperature of the inflection region. We show the three solutions at temperature $T=135\,\mathrm{MeV}$ in the right panel of Fig.\ref{firstorder-PT1}. In principle, we can calculate the free energies corresponding to these solutions to quantify the first-order transition and to fix the critical temperature, as has been done in \cite{Chelabi:2015gpc}. Here it is enough for us to know the order of phase transitions.

\subsection{The three-flavor case}\label{3flavorcase} 

In the three-flavor case ($m_{u}=m_{d}=m_{s}$), there is no distinction between the VEVs $\chi_{u}$ and $\chi_{s}$. Thus the Eqs. (\ref{vevX-eom1}) and (\ref{vevX-eom2}) will reduce to one equation, i.e.,
\begin{align}\label{3flav-vevX-eom1}
\chi'' +\left(\frac{f'}{f}+3A'-\Phi'\right)\chi' -\frac{e^{2A}}{f}\left(m_5^2\chi +\lambda\chi^3 +\frac{\gamma}{2\sqrt{2}}\chi^2\right) &=0 \,,
\end{align}
which has been considered in the previous works (see e.g. \cite{Chelabi:2015cwn,Chelabi:2015gpc}). We can also deal with the two coupled equations (\ref{vevX-eom1}) and (\ref{vevX-eom2}) directly and the result will be the same as that solved from the single equation.

\begin{figure}[h]
\begin{center}
 \includegraphics[width=68mm,clip=true,keepaspectratio=true]{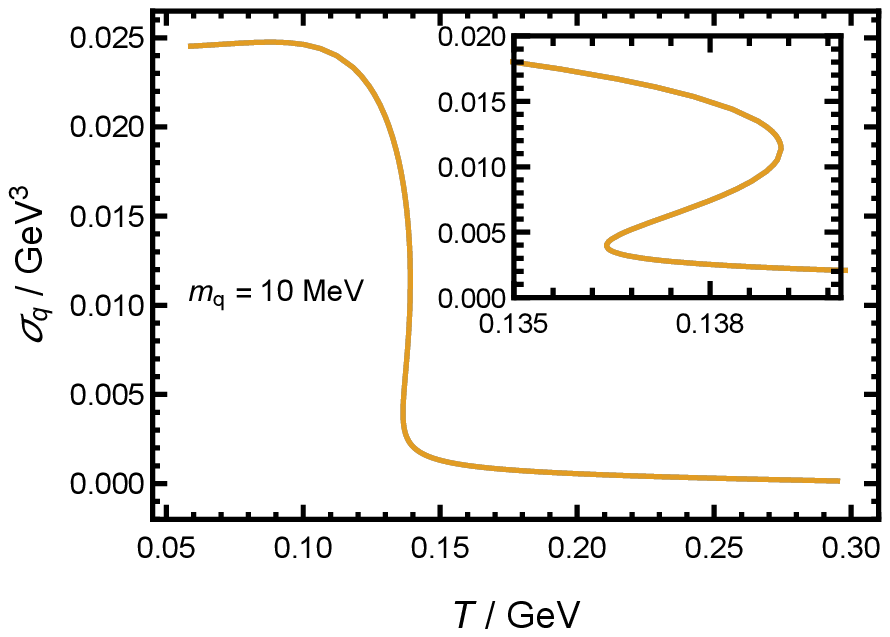}
\hspace*{0.6cm}
\includegraphics[width=68mm,clip=true,keepaspectratio=true]{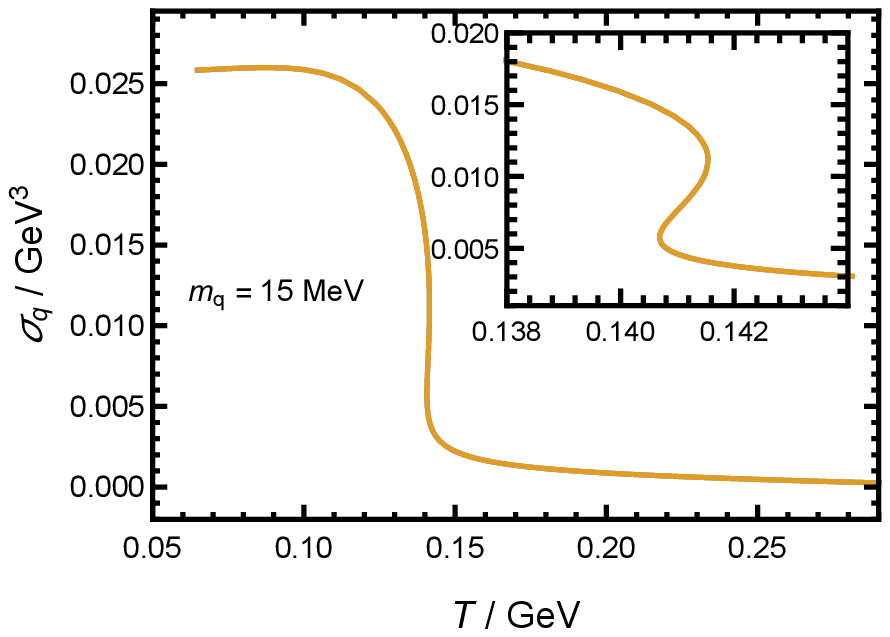}
 \vspace{0.35cm} \\ 
\includegraphics[width=68mm,clip=true,keepaspectratio=true]{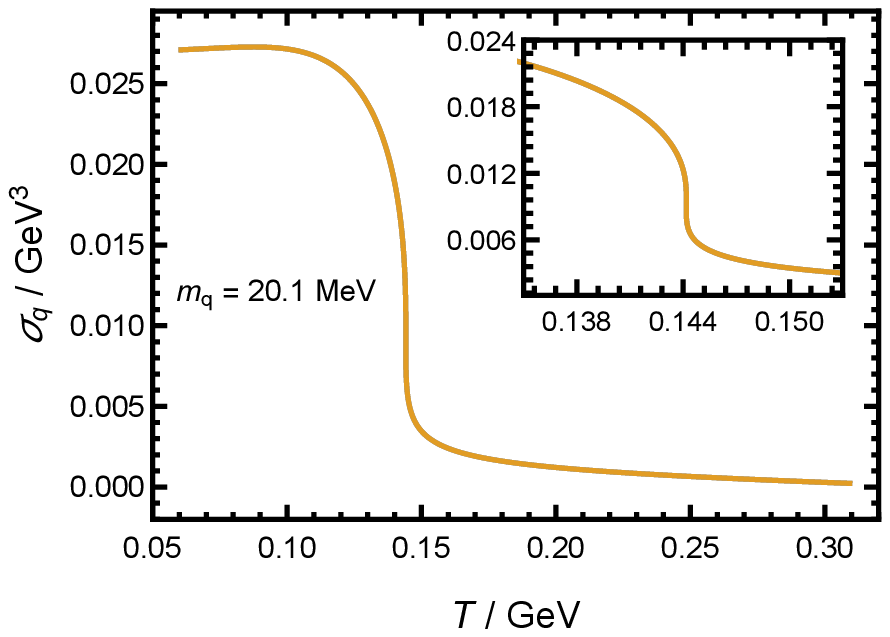}
\hspace*{0.6cm}
\includegraphics[width=68mm,clip=true,keepaspectratio=true]{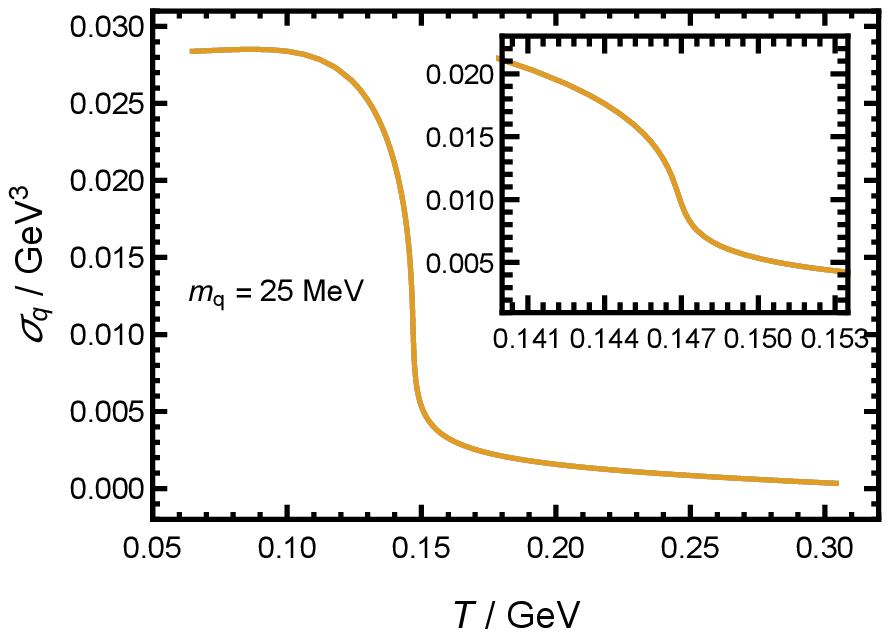}
\vskip -1cm \hskip 0.7 cm
\end{center}
\caption{The chiral transition behavior of the condensate ($\sigma_q\equiv\sigma_u=\sigma_s$) with temperature $T$ at $m_{q}=10, 15, 20.1, 25\,\mathrm{MeV}$ in the three-flavor case ($m_q \equiv m_{u}=m_{d}=m_{s}$).}
\label{chiralPhaseTransition4}
\end{figure}

We present the calculated results of the improved soft-wall AdS/QCD model in Fig.\ref{chiralPhaseTransition4}, where the chiral transition behavior of ($\sigma_{u},\sigma_{s}$) with four different quark masses are plotted. The top two panels show obvious inflections in the transition region, which implies a first-order phase transition. The lower left panel shows a near second-order phase transition, while the lower right one is a crossover transition. We see that at small quark masses the chiral transition is first-order and becomes a second-order one when the quark mass increases up to a point, and then crossover transitions, which is consistent with the phase diagram in Fig.\ref{columbia-plot}. We can also see roughly from Fig.\ref{chiralPhaseTransition4} that the (pseudo-)critical transition temperature increases with increasing quark mass, at least in a finite range of quark masses, which agrees with the lattice result \cite{Laermann:2003cv}.

\subsection{The $m_s=0$ case}\label{ms-0-case1}

\begin{figure}[h]
\begin{center}
 \includegraphics[width=68mm,clip=true,keepaspectratio=true]{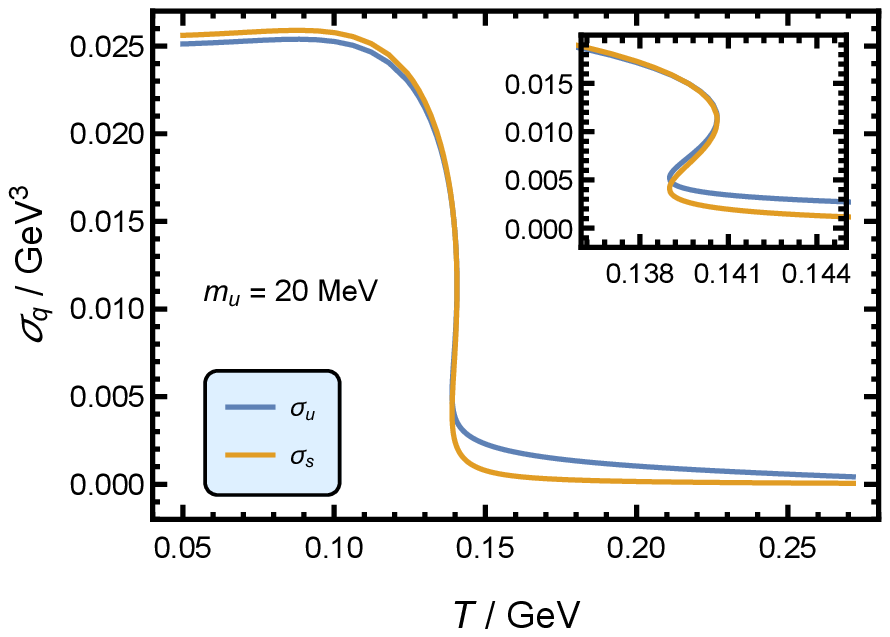}
\hspace*{0.6cm}
\includegraphics[width=68mm,clip=true,keepaspectratio=true]{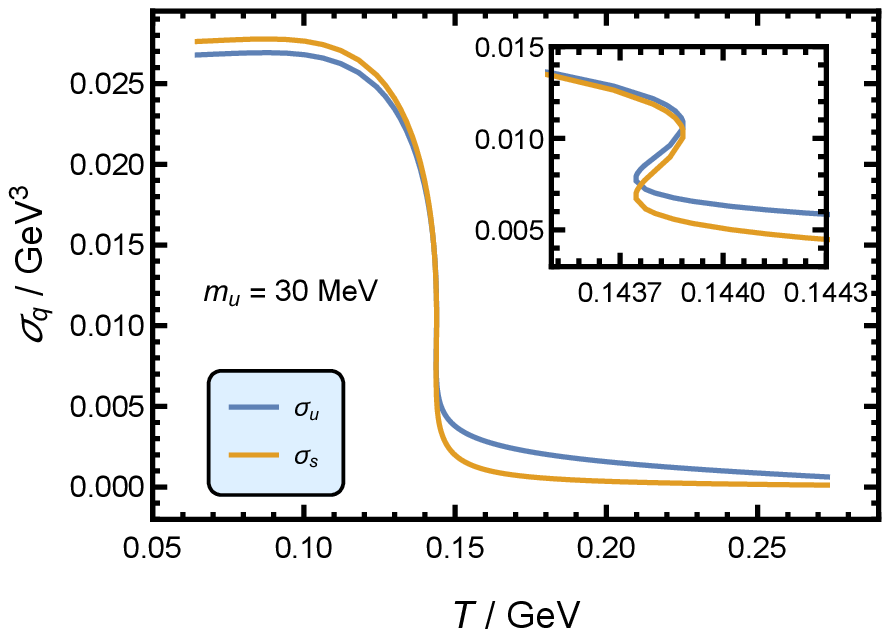}
 \vspace{0.35cm} \\ 
\includegraphics[width=68mm,clip=true,keepaspectratio=true]{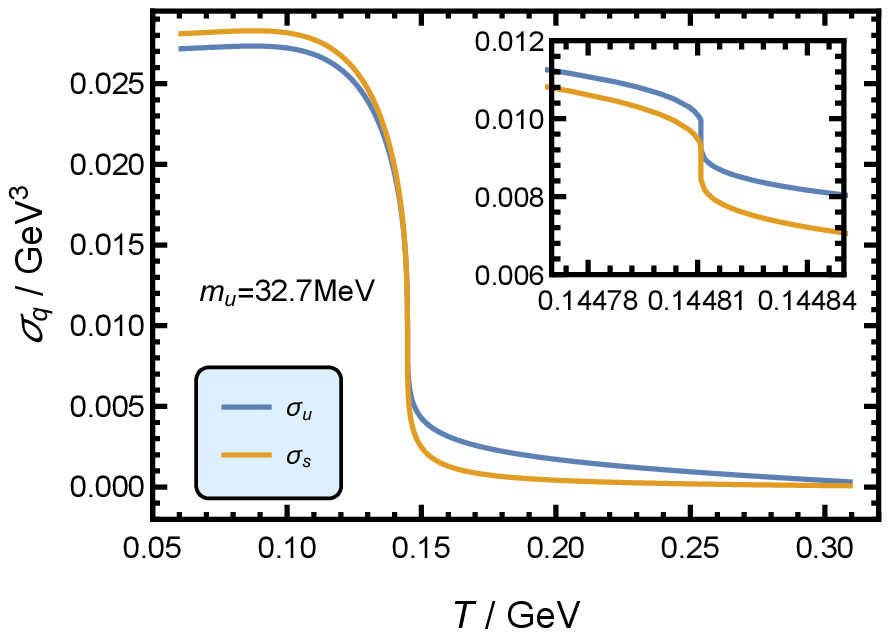}
\hspace*{0.6cm}
\includegraphics[width=68mm,clip=true,keepaspectratio=true]{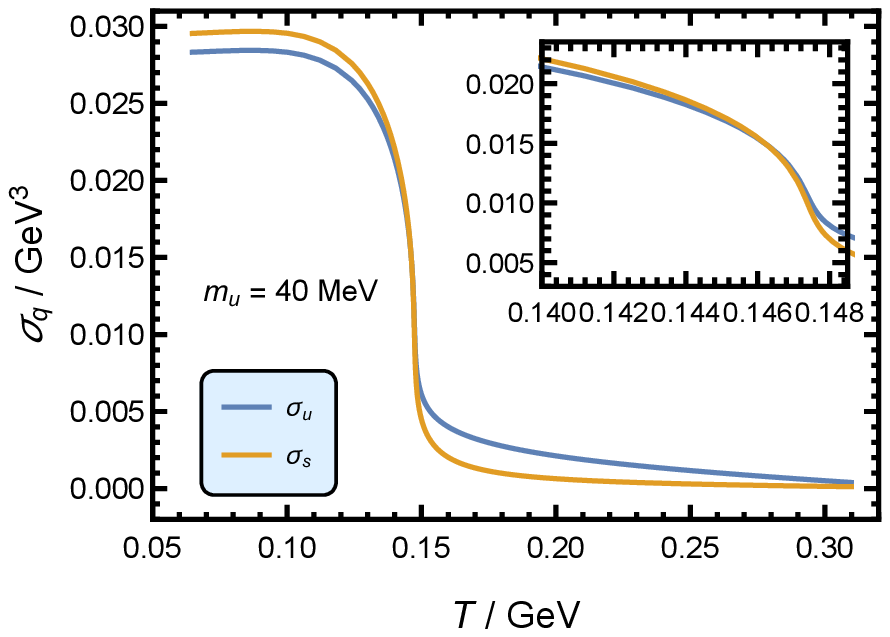}
\vskip -1cm \hskip 0.7 cm
\end{center}
\caption{The chiral transition behavior of the condensates ($\sigma_u$, $\sigma_s$) with temperature $T$ at $m_{u}=20, 30, 32.7, 40\,\mathrm{MeV}$ in the $m_{s}=0$ case.} \label{chiralPhaseTransition3}
\end{figure}

In the $m_{s}=0$ case, we need to solve the two coupled equations (\ref{vevX-eom1}) and (\ref{vevX-eom2}) to obtain the set of solutions ($\chi_{u},\chi_{s}$), from which the condensates ($\sigma_{u},\sigma_{s}$) can be extracted. We plot the chiral transitions of ($\sigma_{u},\sigma_{s}$) with temperature $T$ for four selected $u$ quark masses in Fig.\ref{chiralPhaseTransition3}, from which we can see that with the increasing of the quark mass $m_{u}$ similar transition behavior happens as that in the three-flavor case. When $m_{u}$ is small (see the upper two panels), both $\sigma_{u}$ and $\sigma_{s}$ show a first-order phase transition with an inflection appearing in the critical transition region, and then both of them become second-order transition as $m_{u}$ reaches to an end point around $32.7$ MeV (see the lower left panel). The chiral transitions of ($\sigma_{u},\sigma_{s}$) become crossover when $m_{u}$ is larger than the value of the second-order transition point (see the lower right panel).

Contrary to the case of the physical point in Fig.\ref{chiralPhaseTransitionAndSol1-1}, the condensate $\sigma_{s}$ is larger than $\sigma_{u}$ at low temperatures and smaller than $\sigma_{u}$ at high temperatures in the $m_{s}=0$ case. We find that this is always the case in the lower right region of our calculated phase diagram with $m_{u,d}=m_{s}$ being as the dividing line, as can be expected. In the upper left region, the relationship between $\sigma_{u}$ and $\sigma_{s}$ is reversed, just like the case of the physical point, i.e., $\sigma_{s}$ is smaller than $\sigma_{u}$ at low temperatures and larger than $\sigma_{u}$ at high temperatures. On the dividing line $m_{u,d}=m_{s}$ which corresponds to the three-flavor case, $\sigma_{u}(T)$ and $\sigma_{s}(T)$ merge into a single curve for any quark mass, as shown in Fig.\ref{chiralPhaseTransition4}.

\subsection{The $m_u=0$ case}\label{mu-0-case1}

There are still debates on the order of QCD phase transition for the case of small $m_{u,d}$ but large $m_{s}$ \cite{Laermann:2003cv,Petreczky:2012rq}. The nature of chiral phase transition was first discussed in \cite{Pisarski:1983ms}, where it was argued that the chiral transition should be first-order for three light quark flavors, while for two light quark flavors it could be either second order or first order, which is determined by the fact that to what extent the broken axial symmetry $U_{A}(1)$ is restored at the transition temperature. If the $U_{A}(1)$ symmetry is not restored at the transition temperature, the chiral phase transition in the chiral limit is most likely a second-order one belonging to the $O(4)$ universality class, which is supported by lattice simulations in recent years \cite{Bhattacharya:2014ara}. Indeed, the second-order phase transition already happens at some finite $m_{s}$ (tricritical point) on the $m_u=0$ boundary.

To obtain the chiral transition behavior in the $m_u=0$ case, we need to be careful to solve the Eqs. (\ref{vevX-eom1}) and (\ref{vevX-eom2}). First note that in this case we have two sets of solutions which correspond to $\chi_{u}=0$ and $\chi_{u}\neq 0$ respectively. When $\chi_{u}=0$, Eqs. (\ref{vevX-eom1}) and (\ref{vevX-eom2}) reduce to one equation of $\chi_{s}$
\begin{align}\label{Xu0-eomofvevXs1}
\chi''_{s} +\left(\frac{f'}{f}+3A'-\Phi'\right)\chi'_{s} -\frac{e^{2A}}{f}\left(m_5^2\chi_s +\lambda\chi_s^3\right) &=0 \,,
\end{align}
whose solution will be labeled as $\chi^{0}_{s}$. Thus we have a set of solutions for the Eqs. (\ref{vevX-eom1}) and (\ref{vevX-eom2}), viz. $(\chi_{u},\chi_{s})=(0,\chi^{0}_{s})$. For another set of solutions $(\chi_{u},\chi_{s})$ with $\chi_{u}\neq 0$, we need to solve the two coupled Eqs. (\ref{vevX-eom1}) and (\ref{vevX-eom2}) directly.

\begin{figure}[h]
\begin{center}
 \includegraphics[width=68mm,clip=true,keepaspectratio=true]{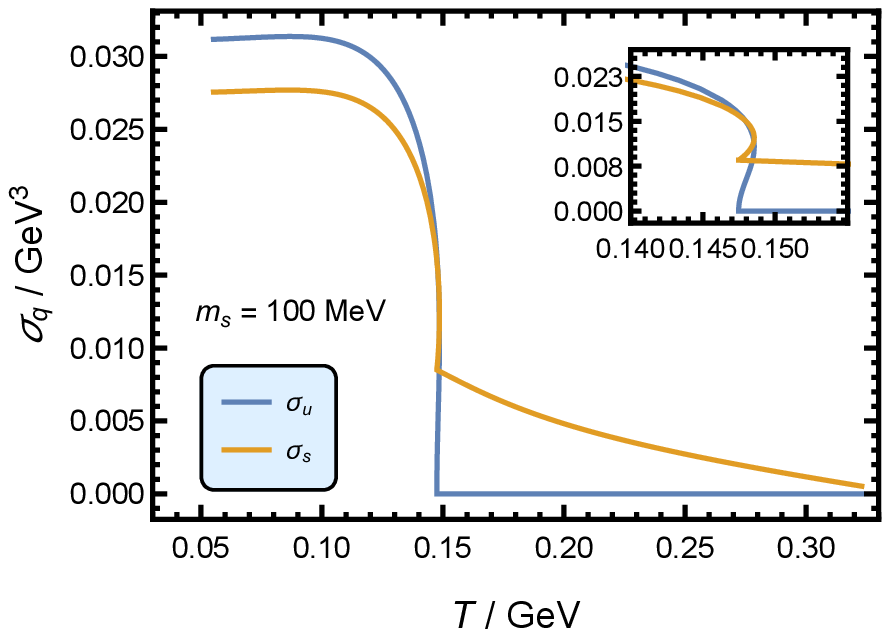}
\hspace*{0.6cm}
\includegraphics[width=68mm,clip=true,keepaspectratio=true]{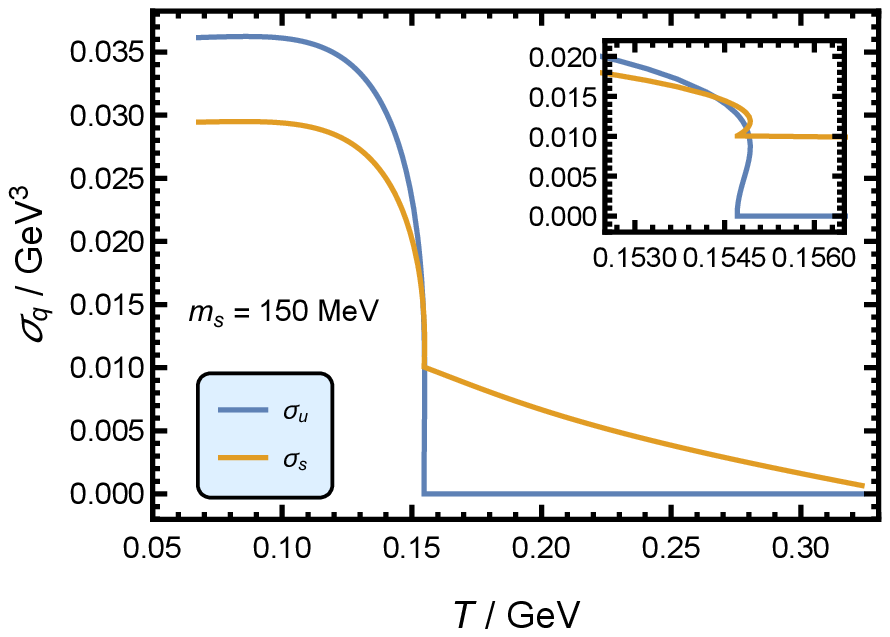}
 \vspace{0.35cm} \\ 
\includegraphics[width=68mm,clip=true,keepaspectratio=true]{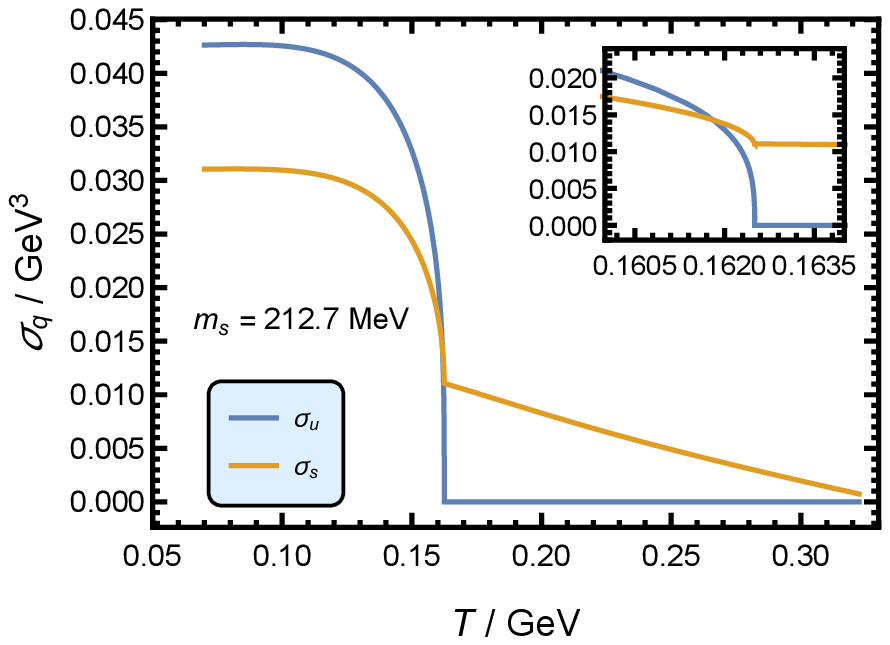}
\hspace*{0.6cm}
\includegraphics[width=68mm,clip=true,keepaspectratio=true]{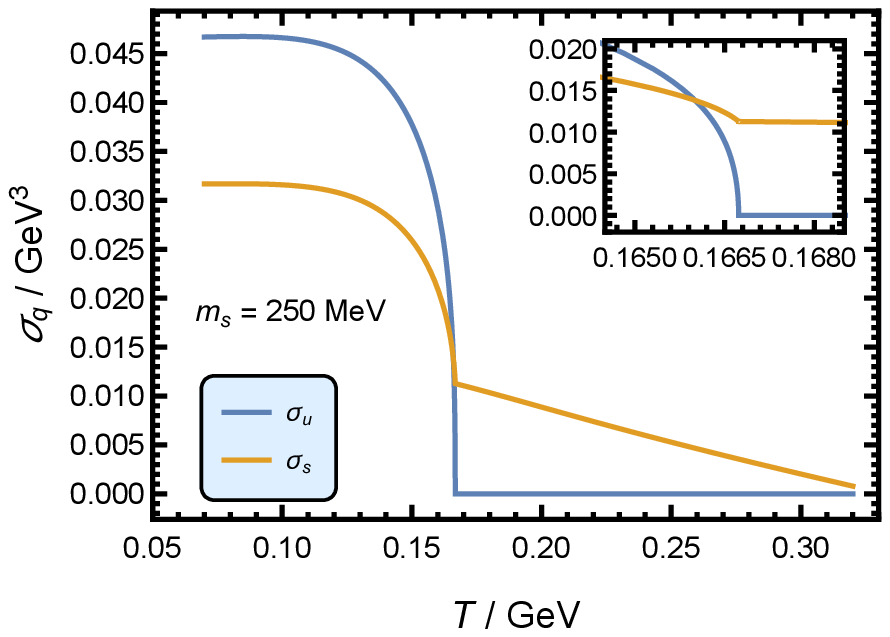}
\vskip -1cm \hskip 0.7 cm
\end{center}
\caption{The chiral transition behavior of the condensates $(\sigma_{u},\sigma_{s})$ with the temperature $T$ at $m_{s}=100, 150, 212.7, 250\,\mathrm{MeV}$ in the $m_{u,d}=0$ case.}
\label{chiralPhaseTransition2}
\end{figure}
The numerical results of the chiral transition behavior of $\sigma_u$ and $\sigma_s$ are shown in Fig.\ref{chiralPhaseTransition2}, where we also choose four typical $s$ quark masses for our consideration. It should be noted that both $\sigma_{u}(T)$ and $\sigma_{s}(T)$ in Fig.\ref{chiralPhaseTransition2} are composed of two parts, of which one corresponds to the solution $(0,\chi^{0}_{s})$ at higher temperatures and the other part corresponds to the solution $(\chi_{u}\neq 0,\chi_{s})$ at lower temperatures. We also note that Eqs. (\ref{vevX-eom1}) and (\ref{vevX-eom2}) have no solution $(\chi_{u}\neq 0,\chi_{s})$ at higher temperatures, and we have dropped the low-temperature part of the solution $(0,\chi^{0}_{s})$ in Fig.\ref{chiralPhaseTransition2} by the stability consideration \cite{Chelabi:2015gpc}. As $\sigma_{u}(T)$ and $\sigma_{s}(T)$ come from two sets of solutions of the Eqs. (\ref{vevX-eom1}) and (\ref{vevX-eom2}), there must be no crossover transition in the $m_u=0$ case. From the upper two panels of Fig.\ref{chiralPhaseTransition2}, we can see obviously the feature of inflection around the transition region, which implies a first-order phase transition at $m_{s}=100, 150\,\mathrm{MeV}$. When $m_{s}$ is large enough, the inflection disappears, which signifies a second-order phase transition (see the lower two panels). The tricritical point, viz. the separation between the first-order and second-order phase transitions on the $m_{u}=0$ boundary is around $m_{s}\simeq212.7\,\mathrm{MeV}$. Thus the results obtained from the improved soft-wall AdS/QCD model support the standard scenario of the ($m_{u,d}, m_s$) phase diagram in Fig.\ref{columbia-plot}.

\section{$\gamma$ dependence of the phase diagram}\label{gamma-depend}

\begin{figure}[h]
\centering
\includegraphics[width=68mm,clip=true,keepaspectratio=true]{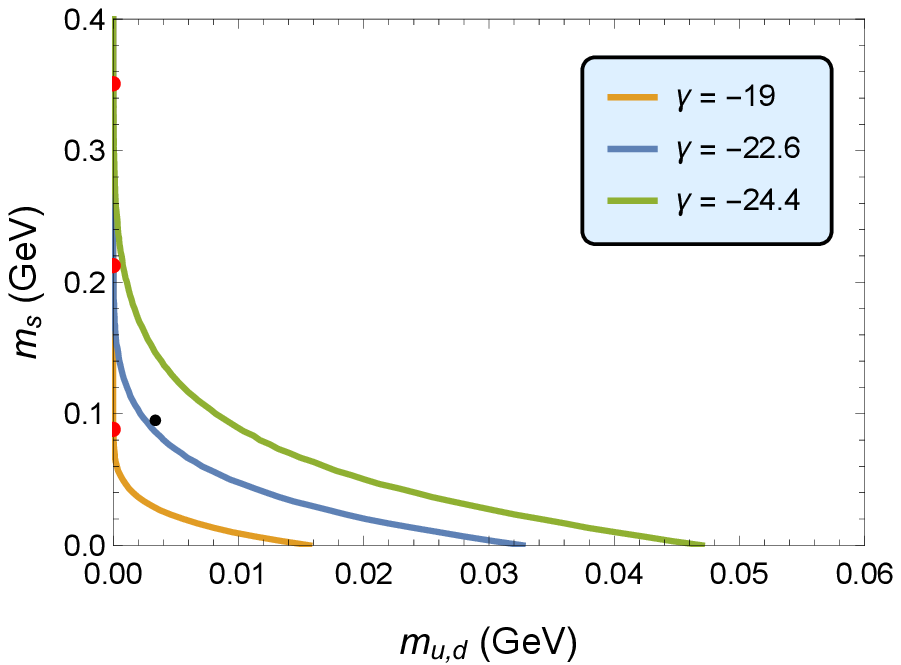}
\hspace*{0.6cm}
\includegraphics[width=69mm,clip=true,keepaspectratio=true]{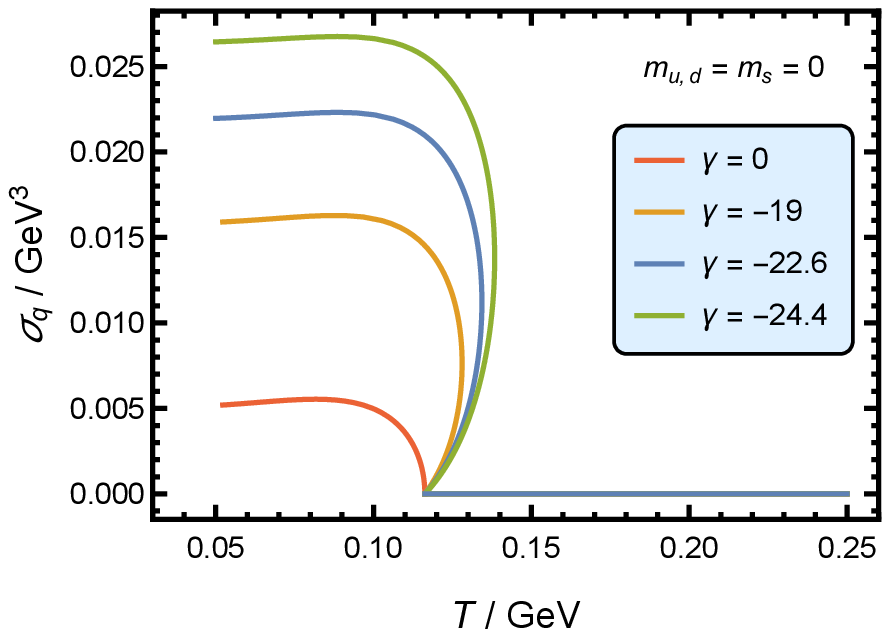}
\vskip -1cm \hskip 0.7 cm
\caption{Left: the calculated ($m_{u,d}, m_s$) phase diagrams in the improved soft-wall model with different values of $\gamma$, where the red dot denotes the tricritical point and the black dot denotes the physical point. Right: the chiral transition behavior of the condensate with temperature $T$ at $m_{u,d}=m_s=0$ for different values of $\gamma$.}
\label{ads-phase-diagram2}
\end{figure}

As mentioned in sec.\ref{parameters1}, the couplings $\gamma$ and $\lambda$ of the improved soft-wall AdS/QCD model are more or less arbitrary in our discussion of chiral phase transition. In this section, we consider the effects of $\gamma$ on the chiral transition behavior and the ($m_{u,d}, m_s$) phase diagram with the coupling $\lambda=16.8$ fixed as before. We choose another two values of $\gamma$ besides the one used in the preceding sections, i.e., $\gamma=-19, -22.6, -24.4$. The second-order transition lines of the phase diagram for the three different values of $\gamma$ are shown in the left panel of Fig.\ref{ads-phase-diagram2}, from which we can see that the second-order line moves upward as $\gamma$ decreases (note that $\gamma\leq 0$). In other words, the region of first-order phase transition would shrink with the increasing of $\gamma$.

In the right panel of Fig.\ref{ads-phase-diagram2}, we plot the chiral transition behavior of the condensate at $m_{u,d}=m_s=0$ when $\gamma$ takes the above three values and also $\gamma=0$. The first-order transition behavior of the condensate is obvious when $\gamma$ takes negative values. However, it becomes a second-order phase transition when $\gamma=0$, which indicates that the region of first-order transition reduces to zero. This can be seen directly from the Eqs. (\ref{vevX-eom1}) and (\ref{vevX-eom2}), which will decouple and reduce to a single equation similar as that in the two-flavor case when the determinant term vanishes ($\gamma=0$). From the right panel of Fig.\ref{ads-phase-diagram2}, we can also see that the condensate at zero or low temperatures will decrease with the increasing of $\gamma$.

\section{Conclusion and discussion}\label{conclusion}

We have studied the chiral phase transition for the $2+1$ quark flavors in the improved soft-wall AdS/QCD model proposed in \cite{Fang:2016nfj}, where we only considered the two-flavor case. This improved soft-wall model can realize consistently the properties of chiral symmetry breaking and linear confinement by a quartic self-coupling term of bulk scalar field and a $z$-dependent bulk scalar mass, which is well motivated by the running quark mass anomalous dimension and the mass split of the chiral partners. The light hadron spectra and many other low-energy quantities have been calculated in \cite{Fang:2016nfj}, which are consistent with experiments. The generalization from the two-flavor case to the $2+1$ flavor case is straight-forward, and the 't Hooft determinant term $\mathrm{det}[X]$ has been included for the generation of correct chiral transition behavior \cite{Chelabi:2015gpc}. The effects of the determinant term on the three-flavor chiral transition in AdS/QCD were first considered in the previous works \cite{Chelabi:2015cwn,Chelabi:2015gpc}, where the model used cannot reproduce the measured hadron spectra.

In this work, we gave the detailed analysis of the chiral transition behavior at different quark masses, and obtained the ($m_{u,d}, m_s$) phase diagram for the quark sector (see Fig.\ref{ads-phase-diagram1}). Quite strikingly, the calculated phase diagram in a simple AdS/QCD model is completely consistent with the standard scenario shown in Fig.\ref{columbia-plot}. It is interesting to find that a tricritical point also exists on the $m_{u}=0$ boundary in the holographic calculation, which is supported by the lattice simulations \cite{Bhattacharya:2014ara}. This tricritical point separates the lower first-order transition region from the upper $O(4)$ second-order line at $m_{u}=0$, which implies that in the chiral limit of two-flavor case the chiral phase transition is second order (consistent with our previous result in \cite{Fang:2016nfj}). However, the reason for the $O(4)$ and $Z(2)$ nature of the second-order line separated by the tricritical point is not very clear in our holographic analysis, which might need more considerations.

In the improved soft-wall AdS/QCD model, there are four parameters relevant to our discussion of chiral phase transition, viz. $\mu_{c}$, $\mu_{g}$, $\gamma$ and $\lambda$. As the parameters $\mu_{c}$ and $\mu_{g}$ are related to the chiral symmetry breaking and $\Lambda_{\mathrm{QCD}}$ energy scales respectively, there are only two relatively arbitrary parameters $\gamma$ and $\lambda$ which might be fixed by the hadron spectrum with $2+1$ quark flavors. With the parameter $\lambda$ fixed, we studied the effects of the parameter $\gamma$ on the $2+1$ chiral transition and the ($m_{u,d}, m_s$) phase diagram. We find that the first-order transition region shrinks with the increasing of $\gamma$ ($\gamma\leq 0$) and finally disappears as $\gamma$ vanishes.

In the light of the good description for chiral phase transition and low-energy hadron physics such as meson spectrum in the improved soft-wall AdS/QCD model, many other works might deserve to be done in this framework. For instance, as we have said, the hadron spectrum with $2+1$ quark flavors can be used to fix the parameters in this model. However, this entails a considerable extension of the model to include the strange quark sector \cite{Sui:2010ay}. We can also study the chemical potential effects and the $\mu-T$ phase diagram by introducing a $U(1)$ gauge field in the model. To characterize QCD phase transition more completely, we should investigate the equation of states and the relevant QCD thermodynamics, which are intimately associated with color deconfining process. The expectation value of Polyakov Loop $\langle L\rangle$ can also be calculated for the deconfining phase transition.

\acknowledgments
The authors are grateful to Y. Tian for the valuable discussions on the numerical method, and would also like to thank D.N. Li for the discussions in the early stage of this work. This work was supported by National Science Foundation of China (NSFC) (11690022, 11475237, 11121064) and Strategic Priority Research Program of the Chinese Academy of Sciences (XDB23030100) as well as the CAS Center for Excellence in Particle Physics (CCEPP), and also partially supported by the China Postdoctoral Science Foundation (2016M601106).

\appendix
\section{A brief introduction to spectral collocation method}
Here we give a brief introduction to the spectral collocation method (see \cite{Shen:2011,Yin:2013} for more details). For a general ordinary differential equation
\begin{equation}\label{generalODE}
\mathcal{N}u(x)=f(x), \quad x \in \boldsymbol{\Omega},
\end{equation}
where $\mathcal{N}$ is a nonlinear operator and $\boldsymbol{\Omega}$ denotes a bounded domain of $\mathbb{R}^1$, we can approximate the solution $u(x)$ of Eq. (\ref{generalODE}) by the sum of a finite sequence of functions:
\begin{equation}\label{approxOfSol1}
u(x) \approx u_N (x) = \sum\limits_{k=0}^{N} a_k \phi_k (x),
\end{equation}
where $\phi_{k}(x)$ are called trial (or basis) functions. Substituting $u_N(x)$ for $u(x)$ in Eq. (\ref{generalODE}) leads to
\begin{equation}\label{residualOfODE}
\boldsymbol{\mathrm{R}}_N (x)=\mathcal{N} u_N(x)-f(x) \not= 0, \quad x \in \boldsymbol{\Omega},
\end{equation}
where $\boldsymbol{\mathrm{R}}_N (x)$ is the residual. We then minimize the residual by requiring
\begin{equation}\label{minResidualIntegral}
\left( \boldsymbol{\mathrm{R}}_N, \psi_j \right)_{\omega} \equiv \int_{\boldsymbol{\Omega}}^{} \boldsymbol{\mathrm{R}}_N (x) \psi_j(x) \omega(x) dx =0, \quad 0 \le j \le N,
\end{equation}
where $\psi_j(x)$ are called test functions and $\omega(x)$ is a positive weight function, or
\begin{equation}\label{minResidualSum}
\langle \boldsymbol{\mathrm{R}}_N, \psi_j \rangle_{N, \omega} \equiv \sum\limits_{k=0}^{N} \boldsymbol{\mathrm{R}}_N (x_k) \psi_j(x_k) \omega_k =0, \quad 0 \le j \le N,
\end{equation}
where $\left\{ x_k \right\}_{k=0}^{N}$ are a set of preselected collocation points and $\left\{ \omega_k \right\}_{k=0}^{N}$ are the weights of a numerical quadrature formula. The above kind of method is called weighted residual method (WRM).
\par
If we employ globally smooth functions as trial/test functions, the WRM is just spectral method. If the test functions $\psi_k (x)$ in (\ref{minResidualSum}) are the Lagrange basis polynomials such that $\psi_j (x_k)=\delta_{j\,k}$ with $\left\{ x_j \right\}$ the preassigned collocation points, then (\ref{minResidualSum}) becomes
\begin{equation}\label{residualInSpectCollocMethod}
\boldsymbol{\mathrm{R}}_N (x_j)=0.
\end{equation}
This kind of choice leads to spectral collocation method.

\section{Implementation of spectral collocation method in our model}
We now detail the implementation of the spectral collocation method in the calculation of chiral phase transition. First we change the form of Eqs. (\ref{vevX-eom1}) and (\ref{vevX-eom2}) by applying the following two transformations successively:
\begin{align}
& z \to u\,z_h, \quad \chi_u (z) \to \hat{\chi}_u (u), \quad \chi_s (z) \to \hat{\chi}_s (u), \label{transfSeqIndex1} \\
& u \to \frac{t+1}{2}, \quad \hat{\chi}_u(u) \to \frac{t+1}{2}\tilde{\chi}_u(t), \quad \hat{\chi}_s(u) \to \frac{t+1}{2}\tilde{\chi}_s(t). \label{transfSeqIndex2}
\end{align}
Thus Eqs. (\ref{vevX-eom1}) and (\ref{vevX-eom2}) are transformed into
\begin{align}
&p_{1}(t)\,\tilde{\chi}_{u}''(t)+p_{2}(t)\,\tilde{\chi}_{u}'(t)+p_{3}(t)\,
\tilde{\chi}_{u}(t)+p_{4}(t)\,\tilde{\chi}_{u}(t)\,\tilde{\chi}_{s}(t)+p_{5}(t)\,
\tilde{\chi}_{u}(t)^{3}=0, \label{vevX-eom3-1} \\
&p_{1}(t)\,\tilde{\chi}_{s}''(t)+p_{2}(t)\,\tilde{\chi}_{s}'(t)+p_{3}(t)\,
\tilde{\chi}_{s}(t)+p_{4}(t)\,\tilde{\chi}_{u}(t)^{2}+p_{5}(t)\,
\tilde{\chi}_{s}(t)^{3}=0, \label{vevX-eom3-2}
\end{align}
where $t \in \left(-1,1\right)$, and the coefficient functions are
\begin{alignat}{3}
\label{coe-vevX-eom3}
&p_{1}(t)=&&-2 \left(-15 - 11\, t + 10\, t^2 + 10\, t^3 + 5\, t^4 + t^5\right), \nonumber&\\
&p_{2}(t)=&&-38 -15\, z_h^2\, \mu_g^2 + 6\, t^5\, z_h^2 \,\mu_g^2 + t^6 \,z_h^2 \,\mu_g^2  -t^2 \left(36 + z_h^2\, \mu_g^2\right)  \nonumber&\\
&&&+ 4\, t^3 \left(-6 + 5 \,z_h^2\,\mu_g^2\right)+3\, t^4 \left(-2 + 5\, z_h^2 \,\mu_g^2\right) - 2\, t \left(12 + 13 \,z_h^2 \,\mu_g^2\right), \nonumber&\\
&p_{3}(t)=&&\left(1 + t\right) \left(-2 + 4\, t^3 \,z_h^2\, \mu_g^2 + t^4 \,z_h^2 \,\mu_g^2 +z_h^2 \,\left(8 \,\mu_c^2 - 15 \,\mu_g^2 \right)\right. &\\
&&&\left.+4\, t \left(-1+ z_h^2 \,\mu_g^2\right) + t^2 \left(-2 + 6 \,z_h^2 \,\mu_g^2\right)\right), \nonumber&\\
&p_{4}(t)=&&-4 \sqrt{2}\, \gamma, \nonumber&\\
&p_{5}(t)=&&-8\left(1 + t\right)\lambda. \nonumber&
\end{alignat}
The UV boundary conditions (\ref{boundCond1}) now become
\begin{align}\label{boundCond3-1}
\tilde{\chi}_u(-1)-m_u\,z_h\,\zeta=0,  \qquad  \tilde{\chi}_s(-1)-m_s\,z_h\,\zeta=0.
\end{align}
As mentioned in sec. \ref{bc-EOM}, the EOMs of the improved soft-wall AdS/QCD model have natural IR boundary conditions, which will be satisfied implicitly in the following procedure.
\par
Now we turn back to the spectral collocation method. Let us first select a set of distinct collocation points $\left\{ t_i \right\}_{i=0}^{N}$ $\left( \text{N is integer and}\; N\ge 1 \right)$ on $\left[ -1, 1 \right]$ with $\left\{ t_i \right\}_{i=0}^{N}$ being the $N+1$ roots of the equation
\begin{equation}\label{EquationOfCollocationPoints}
(1-t^2)\frac{\mathrm{d}}{\mathrm{d}t}L_{N}(t)=0,
\end{equation}
where $L_{N}(t)$ is the $N^{\mathrm{th}}$ Legendre polynomial. The derivative matrix $\mathbf{D}$ is a $\left( N+1 \right) \times \left( N+1 \right)$ matrix which is given by
\begin{align}\label{diffMatIndex1}
\mathbf{D}_{i j}=
\begin{cases}
-\frac{N(N+1)}{4},&i=j=0,\\
0, &1 \le i=j \le N-1,\\
\frac{N(N+1)}{4},&i=j=N,\\
\frac{L_{N}(x_{i})}{(x_{i}-x_{j})L_{N}(x_{j})},&i \neq j.
\end{cases}
\end{align}
Note that the range of each index of the matrix $\mathbf{D}$ is $\left[0,N\right]$. For a general function $f(t)$, $f(t_{i})$ denotes the value of $f(t)$ at the collocation point $t_i$. The value of the derivative function $f'(t)$ can be calculated by the differential matrix:
\begin{equation}\label{diffMatIndex2}
\vv{\mathbf{f'(t)}}=\mathbf{D}\, \vv{\mathbf{f(t)}},
\end{equation}
where
\begin{align}\label{generalFunValVecIndex1}
\vv{\mathbf{f(t)}} =\begin{pmatrix} f(t_0) \\ f(t_1) \\ \vdots \\ f(t_N) \end{pmatrix},  \qquad
\vv{\mathbf{f'(t)}} =\begin{pmatrix} f'(t_0) \\ f'(t_1) \\ \vdots \\ f'(t_N) \end{pmatrix}.
\end{align}
The matrix of second-order differential operator is defined as $\mathbf{D_{2}}=\mathbf{D}^{2}$.
\par
Then we can build the discretization scheme of Eqs. (\ref{vevX-eom3-1}) and (\ref{vevX-eom3-2})
\begin{alignat}{2}
&p_{1}(t_{i})\, \sum_{j=0}^{N}\left(\mathbf{D_{2}}\right)_{ij} \tilde{\chi}_{u}(t_{j})+p_{2}(t_{i})\, \sum_{j=0}^{N}\mathbf{D}_{i j} \tilde{\chi}_{u}(t_{j})+p_{3}(t_{i})\,\tilde{\chi}_{u}(t_{i}) \nonumber & \\
&+p_{4}(t_{i})\, \tilde{\chi}_{u}(t_{i})\tilde{\chi}_{s}(t_{i})+ p_{5}(t_{i})\, \tilde{\chi}_{u}(t_{i})^{3}=0, \quad &1 \le i \le N, \label{discr-vevX-eom3-1} &\\
&p_{1}(t_{i})\, \sum_{j=0}^{N}\left(\mathbf{D_{2}}\right)_{ij} \tilde{\chi}_{s}(t_{j})+p_{2}(t_{i})\, \sum_{j=0}^{N}\mathbf{D}_{i j} \tilde{\chi}_{s}(t_{j})+p_{3}(t_{i})\,\tilde{\chi}_{s}(t_{i}) \nonumber & \\
&+p_{4}(t_{i})\, \tilde{\chi}_{u}(t_{i})^{2}+ p_{5}(t_{i})\, \tilde{\chi}_{s}(t_{i})^{3}=0, \quad &1 \le i \le N.  \label{discr-vevX-eom3-2} &
\end{alignat}
Because the domain of Eqs. (\ref{vevX-eom3-1}) and (\ref{vevX-eom3-2}) is $\left(-1,1\right)$, the range of the index $i$ in Eqs. (\ref{discr-vevX-eom3-1}) and (\ref{discr-vevX-eom3-2}) should be $\left[1,N-1\right]$. However, as the IR boundary conditions are natural boundary conditions which are satisfied implicitly in Eqs. (\ref{discr-vevX-eom3-1}) and (\ref{discr-vevX-eom3-2}), thus we take $1\le i\le N$ with $N$ included. The UV boundary conditions (\ref{boundCond3-1}) now turn into
\begin{align}\label{discr-boundCond3-1}
\tilde{\chi}_{u}(t_{0})-m_u\,z_h\,\zeta=0,  \qquad  \tilde{\chi}_{s}(t_{0})-m_s\,z_h\,\zeta=0.
\end{align}
Putting together Eqs. (\ref{discr-vevX-eom3-1}), (\ref{discr-vevX-eom3-2}) and (\ref{discr-boundCond3-1}), we have totally $2\left( N+1 \right)$ equations which can be solved numerically to obtain the values of $\left(\tilde{\chi}_{u}(t_i), \tilde{\chi}_{s}(t_i)\right)$ $(0\le i \le N)$.



\end{document}